\documentclass[rmp,aps,twocolumn, nofootinbib,a4paper,notitlepage, 14pt]{revtex4}

\usepackage{amsmath}
\usepackage{amssymb}
\usepackage{amsfonts}
\usepackage{epsfig, color, ulem}
\usepackage{graphicx}
\usepackage{tcolorbox}

\usepackage{natbib}
\usepackage{setspace}
\newlength{\dinwidth}
\newlength{\dinmargin}
\usepackage{lineno}

\begin{document}

\begin{spacing}{1}

\title{Representations for multidimensional down-down deconvolution of ocean-bottom seismic data: theory and practical implications.}
\author{\small Kees Wapenaar$^1$, Matteo Ravasi$^2$ and Claudio Bagaini$^3$}
\affiliation{$^1$Delft University of Technology,  Department of Geoscience and Engineering, Stevinweg 1, 2628 CN Delft, The Netherlands. E-mail: c.p.a.wapenaar@tudelft.nl.\\
$^2$Shearwater GeoServices, Gatwick, United Kingdom.\\
$^3$SLB, Italy.\\
\mbox{}\\
{\rm Right-running head: Multidimensional down-down deconvolution}}

\date{\today}

{\small
\begin{abstract}
Multidimensional up-down deconvolution effectively eliminates surface-related multiples from ocean-bottom seismic data.
Recently, several down-down deconvolution methods have been introduced as attractive alternatives.
Whereas multidimensional up-down deconvolution fully accounts for lateral variations of the medium parameters,
the underlying theory of some of the down-down deconvolution methods is essentially based on the assumption that the medium is horizontally layered.
Using reciprocity theory, this assumption is circumvented. 
This leads to representations for either receiver-side or source-side multidimensional down-down deconvolution.
Compared with multidimensional up-down deconvolution, receiver-side down-down deconvolution only utilizes the downgoing part of the wavefield that better samples the shallow subsurface,
but it is not entirely data-driven. Source-side down-down deconvolution benefits from the better sampled source array, but in the presence of sparsely sampled receivers it requires 
solving an underdetermined system of linear equations.
\end{abstract}
\maketitle

\section{Introduction}

Up-down deconvolution is an effective method to eliminate surface-related multiples from ocean-bottom seismic (OBS) data.
An early version assumes that the medium 
is horizontally layered \cite{Sonneland87SEG}.  \citet{Amundsen99SEG} and \citet{Wapenaar2000SEG}
formulate representations for multidimensional up-down deconvolution, which hold for arbitrarily inhomogeneous media.

To tackle illumination issues of shallow reflectors, \citet{Hampson2020EAGE}, \citet{Caprioli2021EAGE} 
and \citet{Lokshtanov2024EAGE} propose down-down deconvolution as an alternative to up-down deconvolution.
Although they apply their methodology to OBS data from laterally varying media, their theory is essentially restricted to horizontally layered media. 
In this paper, we derive a new representation for multidimensional down-down deconvolution. 
This generalizes the theory of the aforementioned authors to the situation of arbitrarily inhomogeneous media.
Since the integrals are along the receiver array, we call this a representation for receiver-side multidimensional  down-down deconvolution.
We compare this with alternative representations for source-side multidimensional  down-down deconvolution, as proposed by \citet{Boiero2023EAGE} and \citet{Wang2024GEO}. 
In the discussion section we discuss the pros and cons of each scheme.

\begin {figure}[t]
\centerline{\hspace{7.5cm}\epsfysize =7.5 cm \epsfbox {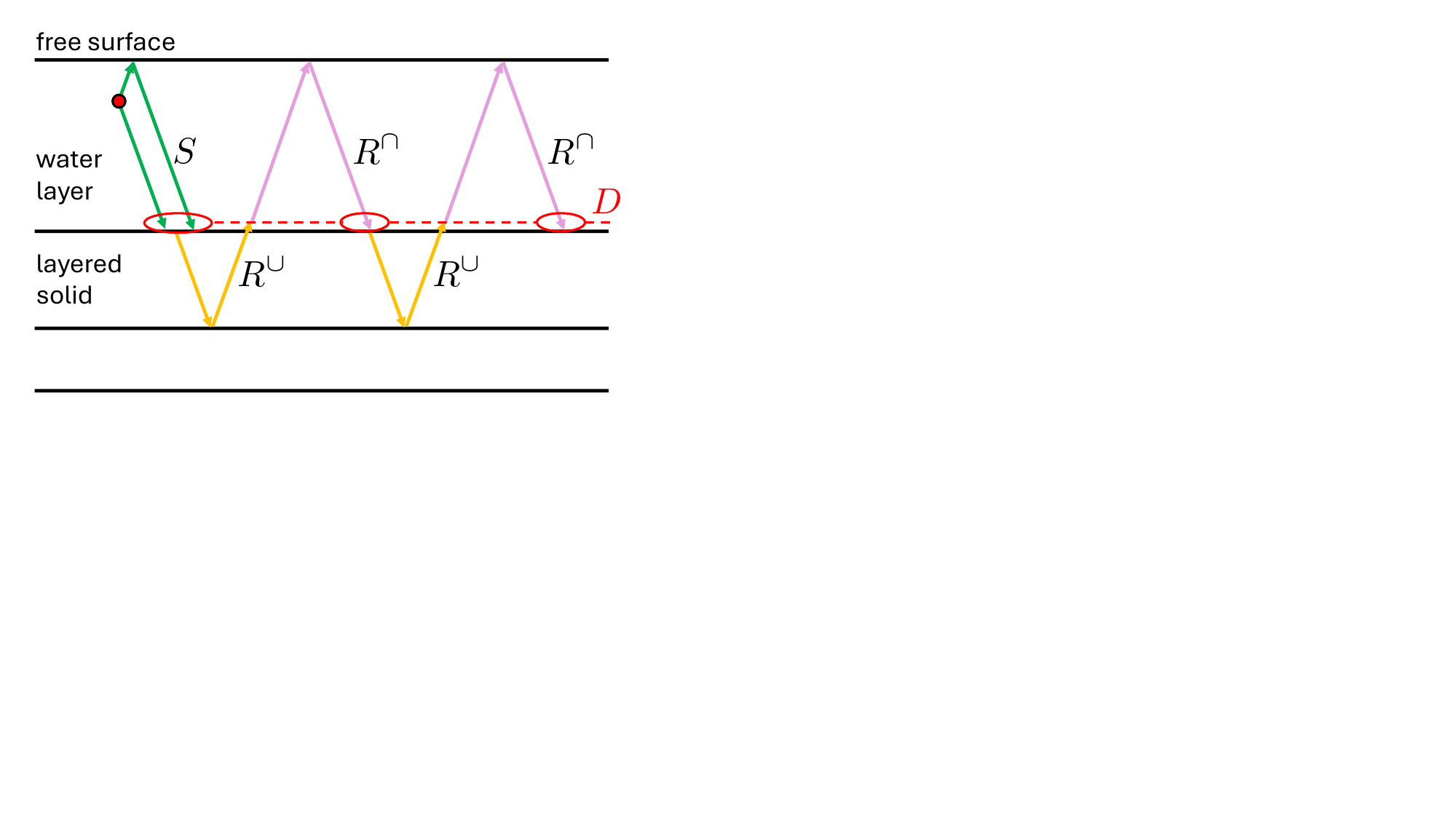}}
\vspace {-4.cm}
\caption {\small Wavefields in horizontally layered OBS configuration (appearing in equations \ref{eq1} -- \ref{eq3}). 
$S$ is the source wavefield (including the free-surface ghost) just above the ocean bottom, and $R^\cap$ is the reflection response of the free surface, seen from just above the ocean bottom.
Both $S$ and $R^\cap$ are defined in the water layer, including the free surface, 
but with the medium below this layer replaced by a homogeneous half-space (we call this state $A$).
$D$ (indicated by the red ellipses)
is the total downgoing field with free-surface multiples, just above the ocean bottom. It is defined in the actual medium, with the free surface (we call this state $B$).
$R^\cup$ is the reflection response of the layered solid medium without free-surface multiples, seen from just above the ocean bottom.
It is defined in the actual medium, but with the free surface replaced by a transparent surface (we call this state $C$).
}\label {Fig1}
\end {figure}

\section{Review of 1-D down-down deconvolution}

We start with a brief review of the down-down deconvolution approach, proposed by \citet{Hampson2020EAGE} and modified by \citet{Caprioli2021EAGE}  and
\citet{Lokshtanov2024EAGE} for the situation of OBS data.
Figure \ref{Fig1} schematically shows the wavefields in a horizontally layered OBS configuration. 
Here, $S$ is the source wavefield (including the free-surface ghost), reaching the ocean bottom from above,
$R^\cup$ is the reflection response of the layered solid medium (without free-surface reflections)
for sources and receivers just above the ocean bottom, and $R^\cap$ is the reflectivity of the free surface, including the 
propagation paths through the water layer. In the rayparameter-frequency domain, the total downgoing wavefield $D$  just above the ocean bottom 
(indicated by the red ellipses in Figure \ref{Fig1}) is given by
\begin{eqnarray}
D&=&S+R^\cap R^\cup S+(R^\cap R^\cup)^2 S+\cdots\nonumber\\
&=&(1-R^\cap R^\cup)^{-1}S.\label{eq1}
\end{eqnarray}
In the rayparameter-frequency domain, all quantities in this equation are scalar functions. Hence, the products are simple scalar products, corresponding to temporal convolutions
(per rayparameter) in the rayparameter-time domain.

With some simple manipulations, equation \ref{eq1}  can be rewritten as
\begin{eqnarray}
D-S=R^\cap R^\cup D.\label{eq2}
\end{eqnarray}
We are interested in retrieving the reflection response $R^\cup$ of the solid medium (i.e., without free-surface reflections).
Assuming the acoustic pressure and the vertical component of the particle velocity are measured at the ocean bottom, the downgoing wavefield $D$ just above the ocean bottom can be obtained by
acoustic decomposition. Assuming the source and the parameters of the water layer are known, $S$ and $R^\cap$ can be obtained by numerical modeling. With these quantities given, 
$R^\cup$ follows from $R^\cup=(D-S)/R^\cap D$. Hence, in the rayparameter-frequency 
domain, $R^\cup$ is retrieved by dividing the muted downgoing field $D-S$ by the scaled downgoing field $R^\cap D$.
Alternatively, $R^\cap R^\cup$ follows from $R^\cap R^\cup=(D-S)/D$.
Both approaches correspond to 1-D down-down deconvolution per rayparameter (i.e., per plane wave) in the rayparameter-time domain. 

\section{Representation for receiver-side down-down deconvolution}

The 1-D deconvolution approach breaks down when the  medium (above and/or below the ocean bottom) is laterally varying.
For this situation, we need a multidimensional version of equation \ref{eq2}, as a basis for multidimensional down-down deconvolution.
Similar to multidimensional up-down deconvolution, we derive  the new multidimensional representation for down-down deconvolution from reciprocity theorems. 
In general, a reciprocity theorem formulates a relation between wavefields in two different states. A complication for the generalization of equation \ref{eq2} is that the 
wavefields in this equation are defined in three different states (see Figure \ref{Fig1}): 
$S$ and $R^\cap$ are defined in the water layer with the free surface and a homogeneous half-space below this layer (we call this state $A$),
$D$ is defined in the actual medium with the free surface (state $B$), 
and $R^\cup$ is defined in the actual medium without the free surface (state $C$).
To deal with these three states, we rewrite equation \ref{eq2} as follows
\begin{eqnarray}
D-S=R^\cap U,\quad\mbox{with}\quad U=R^\cup D,\label{eq3}
\end{eqnarray}
where $U$ is the total upgoing wavefield just above the ocean bottom in the actual medium with the free surface (state $B$).
The first of these equations contains wavefields in states $A$ and $B$, the second in states $B$ and $C$. 
We separately derive the multidimensional versions
of these two equations, which in the end we combine into a multidimensional version of equation \ref{eq2}.

Consider a spatial domain $\mathbb{D}_I$, enclosed by boundaries $\partial\mathbb{D}_0$ (the free surface) and $\partial\mathbb{D}_R$ (just above the ocean bottom),
with outward pointing normal vectors ${\bf n}=(n_1,n_2,n_3)$, see states $A$ and $B$ in Figure \ref{Fig2}. 
Boundary $\partial\mathbb{D}_0$ is not necessarily horizontal (to account for a possibly rough sea surface).
Note that $\mathbb{D}_I$ corresponds with the  water column.
In the space-frequency $({\bf x},\omega)$ domain, the reciprocity theorem for the acoustic wavefields in states $A$ and $B$ reads \cite{Fokkema93Book}
\begin{eqnarray}\label{eq4}
\oint_{\partial\mathbb{D}}\bigl( p_A v_{k,B} - v_{k,A} p_B\bigr)n_k{\rm d}^2{\bf x}
=\int_{\mathbb{D}_I}\bigl( p_A q_B- q_A p_B\bigr){\rm d}^3{\bf x},
\end{eqnarray}
where $p({\bf x},\omega)$ is the acoustic pressure, $v_k({\bf x},\omega)$ for $k=1,2,3$ the particle velocity, $q({\bf x},\omega)$ the volume injection-rate density, and
$\partial\mathbb{D}=\partial\mathbb{D}_0\cup\partial\mathbb{D}_R$ (the contribution of the integral over a cylindrical side boundary with infinite radius vanishes). 
Einstein's summation convention applies to the repeated subscript $k$.
Since the pressure vanishes at the free surface $\partial\mathbb{D}_0$ in both states, the boundary integral can be restricted to
$\partial\mathbb{D}_R$, with $n_1=n_2=0$ and  $n_3=+1$ (we assume the positive $x_3$-axis is pointing downward). 
Applying pressure-normalized up-down decomposition at $\partial\mathbb{D}_R$, we obtain \cite[Appendix B]{Wapenaar89Book}
\begin{eqnarray}\label{eq5}
&&-\frac{2}{i\omega\rho}\int_{\partial\mathbb{D}_R}\bigl((\partial_3 p_A^+) p_B^- + (\partial_3 p_A^-) p_B^+\bigr){\rm d}^2{\bf x}\nonumber\\
&&\hspace{1cm}=\int_{\mathbb{D}_I}(p_A q_B-q_A p_B){\rm d}^3{\bf x},
\end{eqnarray}
where  superscripts $+$ and $-$ denote downward and upward propagation, respectively, $\rho$ is the mass density of the  water column, 
$i$ is the imaginary unit, and $\partial_3$ stands for differentiation in the $x_3$-direction. 
Pressure-normalized decomposition implies $p_A^++p_A^-=p_A$ and $p_B^++p_B^-=p_B$.
Note that the derivation of equation \ref{eq5} from equation \ref{eq4} relies on the assumption that $\partial\mathbb{D}_R$ is horizontal.
The derivation can be extended for flux-normalized decomposed fields at a curved boundary \cite{Frijlink2010SIIMS}, but this is beyond the scope of this paper.

\begin {figure}[t]
\centerline{\hspace{6.5cm}\epsfysize =8 cm \epsfbox {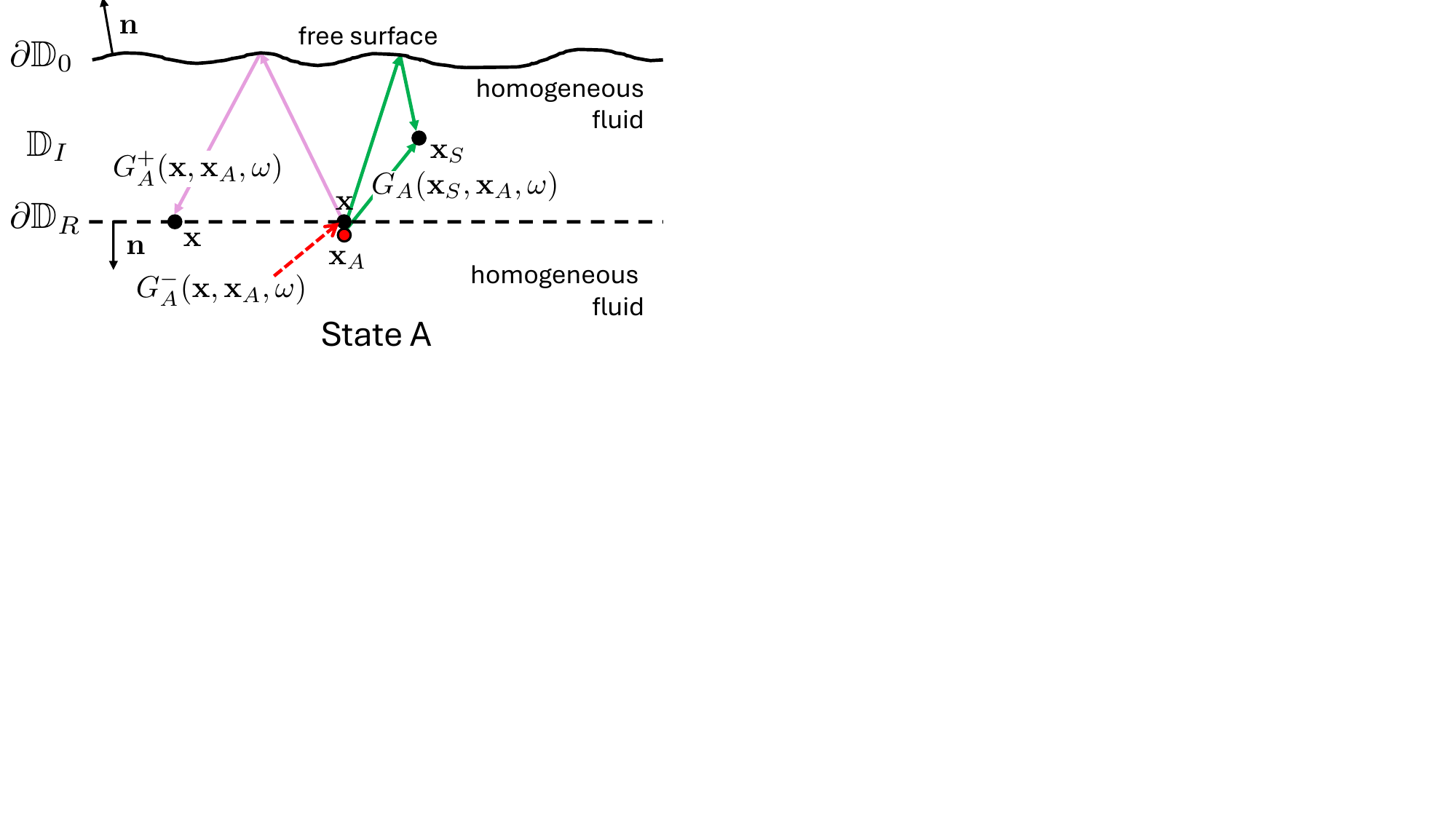}}
\vspace {-3.8cm}
\centerline{\hspace{6.5cm}\epsfysize =8 cm \epsfbox {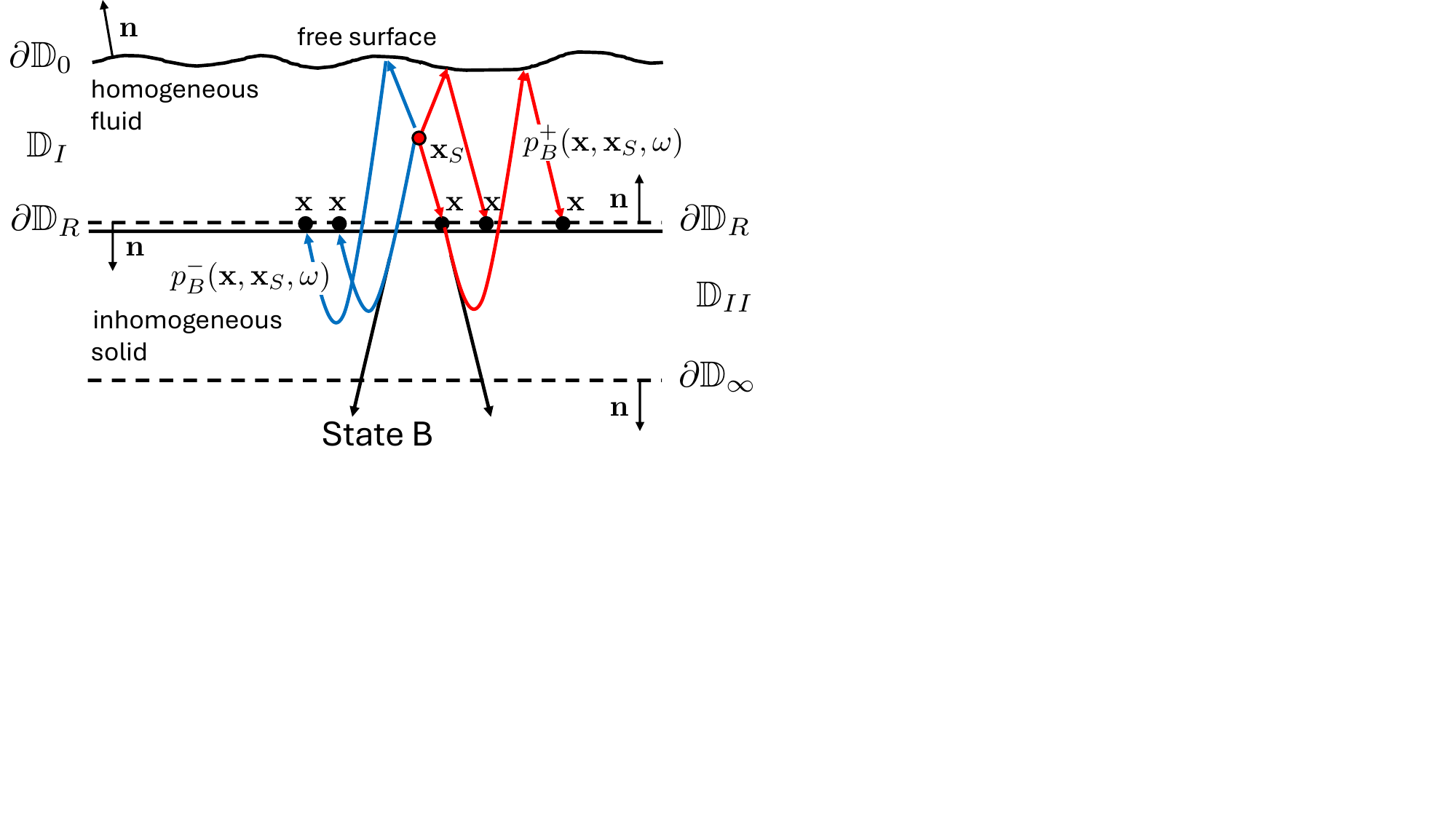}}
\vspace {-2.8cm}
\centerline{\hspace{6.5cm}\epsfysize =8 cm \epsfbox {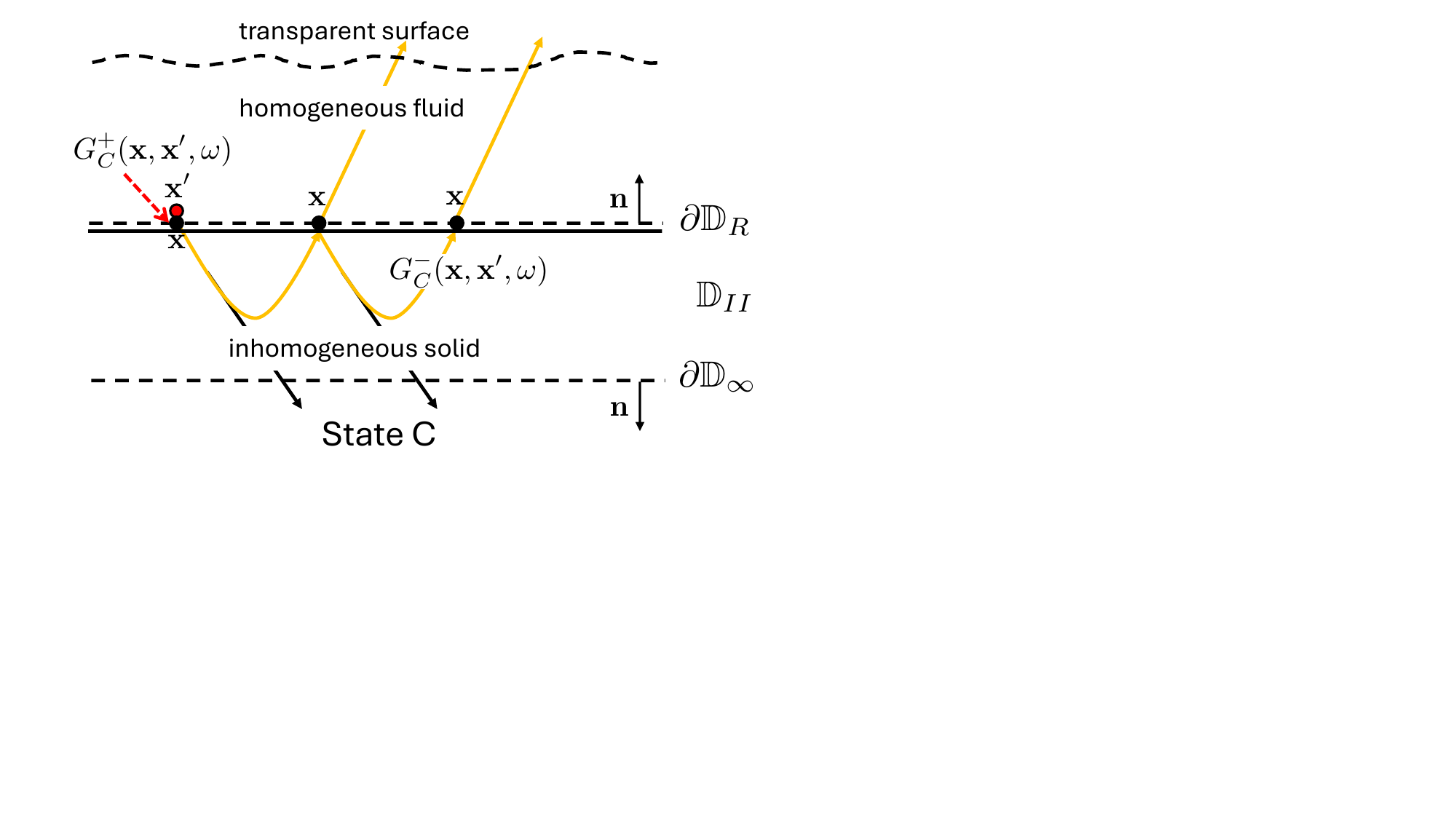}}
\vspace {-3.2cm}
\caption {\small Green's functions,  used for the derivation of the receiver-side multidimensional  down-down deconvolution
representation (equation \ref{eq9}), in an arbitrarily inhomogeneous  medium. In state $A$ (the water layer with the free surface and a homogeneous half-space below this layer), 
the Green's function $G_A({\bf x}_S,{\bf x}_A,\omega)=G_A({\bf x}_A,{\bf x}_S,\omega)$ replaces $S$ and
$G_A^+({\bf x},{\bf x}_A,\omega)$ replaces $R^\cap$ in Figure \ref{Fig1}.
In state $B$ (the actual medium with the free surface), the downgoing wavefield $p_B^+({\bf x},{\bf x}_S,\omega)$
replaces $D$  in Figure \ref{Fig1} (and $p_B^-({\bf x},{\bf x}_S,\omega)$ replaces $U$ in equation \ref{eq3}).
In state $C$ (the actual medium without the free surface), the Green's function $G_C^-({\bf x},{\bf x}',\omega)$
replaces $R^\cup$ in Figure \ref{Fig1}.
}\label {Fig2}
\end {figure}

With reference to Figure \ref{Fig2}, for state $B$, we choose the actual medium, and the wavefield in this state is the response to a  source 
$q_B({\bf x},\omega)=\delta({\bf x}-{\bf x}_S)s(\omega)$, with ${\bf x}_S$ in $\mathbb{D}_I$ and where $s(\omega)$ is the source signature. 
Hence, for ${\bf x}$ in $\mathbb{D}_I$ we have $p_B({\bf x},{\bf x}_S,\omega)=G_B({\bf x},{\bf x}_S,\omega)s(\omega)$,
where $G_B({\bf x},{\bf x}_S,\omega)$ is the Green's function in state $B$. Here and in the following, the second coordinate vector  refers to a source position and the first to a receiver position. 
For ${\bf x}$ at $\partial\mathbb{D}_R$ we express $p_B^\pm$ in terms of a pressure-normalized decomposed Green's function, according to
 $p_B^\pm({\bf x},{\bf x}_S,\omega)=G_B^\pm({\bf x},{\bf x}_S,\omega)s(\omega)$, where 
superscripts $+$ and $-$ denote the downgoing and upgoing components of the wavefield at the receiver position ${\bf x}$. 
For state $A$, we replace the half-space below $\partial\mathbb{D}_R$
by a homogeneous fluid with the same properties as the fluid above $\partial\mathbb{D}_R$, such that the entire medium below the free surface is homogeneous.
 We choose a unit source $q_A({\bf x},\omega)=\delta({\bf x}-{\bf x}_A)$, with ${\bf x}_A$ just below $\partial\mathbb{D}_R$,
i.e., outside $\mathbb{D}_I$. Then for ${\bf x}$ in $\mathbb{D}_I$ we have $p_A({\bf x},{\bf x}_A,\omega)=G_A({\bf x},{\bf x}_A,\omega)$ and for ${\bf x}$ at $\partial\mathbb{D}_R$ we 
consider the decomposed field
$p_A^\pm({\bf x},{\bf x}_A,\omega)=G_A^\pm({\bf x},{\bf x}_A,\omega)$. The upgoing field for ${\bf x}$ at $\partial\mathbb{D}_R$ is non-zero only vertically above ${\bf x}_A$, according to
$\partial_3p_A^-({\bf x},{\bf x}_A,\omega)=\partial_3G_A^-({\bf x},{\bf x}_A,\omega)=-\frac{1}{2}i\omega\rho\delta({\bf x}_{\rm H}-{\bf x}_{{\rm H},A})$ 
\cite[Appendix A]{Wapenaar2017GP2}, where
${\bf x}_{\rm H}$ and ${\bf x}_{{\rm H},A}$ denote the horizontal coordinates of ${\bf x}$ and ${\bf x}_A$, respectively.
The downgoing field for ${\bf x}$ at $\partial\mathbb{D}_R$ is the ``reflection response from below'' of the fluid layer  above $\partial\mathbb{D}_R$ (bounded by the free surface), according to 
$\partial_3 p_A^+({\bf x},{\bf x}_A,\omega)=\partial_3 G_A^+({\bf x},{\bf x}_A,\omega)=\frac{1}{2}i\omega\rho R^\cap({\bf x}_A,{\bf x},\omega)$ 
(note that due to the operator $\partial_3$ acting on ${\bf x}$ and the interchangement of ${\bf x}$ and ${\bf x}_A$, 
 the source at ${\bf x}$ in $ R^\cap({\bf x}_A,{\bf x},\omega)$ is a dipole source). Substituting these choices into equation \ref{eq5},
using $G_A({\bf x}_S,{\bf x}_A,\omega)=G_A({\bf x}_A,{\bf x}_S,\omega)=G_A^+({\bf x}_A,{\bf x}_S,\omega)$, yields
\begin{eqnarray}\label{eq6}
&&\hspace{-1cm}p_B^+({\bf x}_A,{\bf x}_S,\omega)- G_A^+({\bf x}_A,{\bf x}_S,\omega)s(\omega)=\nonumber\\
&&\hspace{0.5cm}\int_{\partial\mathbb{D}_R} R^\cap({\bf x}_A,{\bf x},\omega) p_B^-({\bf x},{\bf x}_S,\omega){\rm d}^2{\bf x}.
\end{eqnarray}
This new representation is, term by term, the multidimensional extension of the first expression in equation \ref{eq3}.

Next, we consider domain $\mathbb{D}_{II}$, enclosed by horizontal boundaries $\partial\mathbb{D}_R$ (just above the ocean bottom) and  $\partial\mathbb{D}_\infty$ 
(below all inhomogeneities in the solid), see states $B$ and $C$ in Figure \ref{Fig2}. 
Hence, $\mathbb{D}_{II}$ consists of a thin fluid domain $\mathbb{D}_f$ (between $\partial\mathbb{D}_R$ and the ocean bottom) and a solid domain $\mathbb{D}_s$
(between the ocean bottom and $\partial\mathbb{D}_\infty$), i.e., $\mathbb{D}_{II}=\mathbb{D}_f\cup\mathbb{D}_s$. 
This time we start with the reciprocity theorem for a fluid-solid configuration of equation A--1 (with subscripts $A$ replaced by $C$).
Since all elastic wavefields at $\partial\mathbb{D}_\infty$ (which corresponds to $\partial\mathbb{D}_s$ in equation A--1) 
are outgoing, the boundary integral over $\partial\mathbb{D}_\infty$ vanishes \cite{Pao76JASA}. 
If we further assume that $\mathbb{D}_{II}$ is source-free in both states, 
we are left with the boundary integral over $\partial\mathbb{D}_R$ (which corresponds to $\partial\mathbb{D}_f$ in equation A--1) being equal to zero. 
After up-down decomposition (similar as in equation \ref{eq5}) we thus obtain
\begin{eqnarray}\label{eq7}
\frac{2}{i\omega\rho}\int_{\partial\mathbb{D}_R}\bigl((\partial_3 p_C^+) p_B^- + (\partial_3 p_C^-) p_B^+\bigr){\rm d}^2{\bf x}=0.
\end{eqnarray}
With reference to Figure \ref{Fig2}, for state $C$, we choose the actual medium, with the free surface replaced by a transparent surface. 
The wavefield in this state is the response to a unit source 
$q_C({\bf x},\omega)=\delta({\bf x}-{\bf x}')$, with ${\bf x}'$ just above $\partial\mathbb{D}_R$, i.e., outside $\mathbb{D}_{II}$.
For  ${\bf x}$ at $\partial\mathbb{D}_R$ we express $ p_C^\pm$ as a decomposed Green's function, according to 
$ p_C^\pm({\bf x},{\bf x}',\omega)= G_C^\pm({\bf x},{\bf x}',\omega)$. The downgoing field for ${\bf x}$ at $\partial\mathbb{D}_R$ is non-zero only vertically below ${\bf x}'$, according to
$\partial_3 p_C^+({\bf x},{\bf x}',\omega)=\partial_3 G_C^+({\bf x},{\bf x}',\omega)=\frac{1}{2}i\omega\rho\delta({\bf x}_{\rm H}-{\bf x}_{{\rm H}}')$, where
${\bf x}_{{\rm H}}'$ denotes the horizontal coordinates of ${\bf x}'$.
The upgoing field for ${\bf x}$ at $\partial\mathbb{D}_R$ is the reflection response  of the fluid-solid configuration below $\partial\mathbb{D}_R$, according to 
$\partial_3 p_C^-({\bf x},{\bf x}',\omega)=\partial_3 G_C^-({\bf x},{\bf x}',\omega)=-\frac{1}{2}i\omega\rho R^\cup({\bf x}',{\bf x},\omega)$.
For state $B$ we make the same choices as before, noting that  ${\bf x}_S$ is outside $\mathbb{D}_{II}$. 
Substituting these choices into equation \ref{eq7} yields
\begin{eqnarray}\label{eq8}
 p_B^-({\bf x}',{\bf x}_S,\omega)=\int_{\partial\mathbb{D}_R} R^\cup({\bf x}',{\bf x},\omega) p_B^+({\bf x},{\bf x}_S,\omega){\rm d}^2{\bf x}.
\end{eqnarray}
This is the multidimensional extension of the second expression in equation \ref{eq3}, and it coincides with equation 7 in \citet{Amundsen2001GEOb}
(or equation 17 in \citet{Wapenaar2000SEG}). Substituting equation \ref{eq8} 
into equation \ref{eq6} (with ${\bf x}$ replaced by ${\bf x}'$) yields
\begin{eqnarray}
&&\hspace{-.5cm}p_B^+({\bf x}_A,{\bf x}_S,\omega)- G_A^+({\bf x}_A,{\bf x}_S,\omega)s(\omega)=\label{eq9}\\
&&\hspace{-.5cm} \int_{\partial\mathbb{D}_R}\int_{\partial\mathbb{D}_R} 
R^\cap({\bf x}_A,{\bf x}',\omega) R^\cup({\bf x}',{\bf x},\omega) p_B^+({\bf x},{\bf x}_S,\omega){\rm d}^2{\bf x}{\rm d}^2{\bf x}'.\nonumber
\end{eqnarray}
This is, term by term, the multidimensional extension of equation \ref{eq2}, where $p_B^+({\bf x},{\bf x}_S,\omega)$ stands for the measured total downgoing wavefield
and $G_A^+({\bf x}_A,{\bf x}_S,\omega)s(\omega)$ for the modeled downgoing source wavefield
(including the free-surface ghost) at $\partial\mathbb{D}_R$, just above the ocean bottom.
This new representation forms a basis for multidimensional down-down deconvolution, aiming to retrieve
the reflection response $R^\cup$ at $\partial\mathbb{D}_R$ of the inhomogeneous solid, without free-surface multiples.
This generalizes existing 1-D down-down deconvolution approaches based on equation \ref{eq2}.
In practice, all quantities in equation \ref{eq9} are discretized in space, and the integrals become matrix products. Hence, 
in practice multidimensional down-down deconvolution involves matrix inversion per frequency component.
Pros and cons relative to other multidimensional deconvolution methods are discussed in the discussion section.

\section{Representation for source-side down-down deconvolution}

The representation of equation \ref{eq9} contains integrals over receivers at $\partial\mathbb{D}_R$. Since receivers are often sparsely sampled in OBS acquisition,  
\citet{Boiero2023EAGE} and \citet{Wang2024GEO} derive alternative representations for down-down deconvolution, containing an integral over the sources.
Here, we derive such a representation, using the formalism of the previous section, to facilitate the comparison between the two approaches.

Consider a domain $\mathbb{D}_f\cup\mathbb{D}_s$ (where subscripts $f$ and $s$ stand for fluid and solid, respectively), 
enclosed by horizontal boundaries  $\partial\mathbb{D}_S$ (just below the sources in the fluid) and $\partial\mathbb{D}_\infty$
(below all inhomogeneities in the solid), see Figure \ref{Fig3}. 
Since in the following derivations we don't encounter integrals along the receivers at the ocean bottom, there is no need to assume that
the ocean bottom (the boundary between $\mathbb{D}_f$ and $\mathbb{D}_s$) is horizontal.
We use again the reciprocity theorem of equation A--1 for the fluid-solid configuration.
As before, the boundary integral over $\partial\mathbb{D}_\infty$ (which corresponds to $\partial\mathbb{D}_s$ in equation A--1) vanishes.
Assuming the solid domain $\mathbb{D}_s$ is source-free, we are left with a boundary integral over $\partial\mathbb{D}_S$ 
(which corresponds to $\partial\mathbb{D}_f$ in equation A--1) and a domain integral over  $\mathbb{D}_f$.
Applying pressure-normalized up-down decomposition, this time not only to the wavefields but also to the sources \cite{Wapenaar2020AMP}, we thus have
\begin{eqnarray}
&&\hspace{-1cm}\frac{2}{i\omega\rho}\int_{\partial\mathbb{D}_S}\bigl((\partial_3 p_A^+) p_B^- + (\partial_3 p_A^-) p_B^+\bigr){\rm d}^2{\bf x}=\nonumber\\
&&\int_{\mathbb{D}_f}\bigl( p_A^+ q_B^-+ p_A^- q_B^+- q_A^+ p_B^-- q_A^- p_B^+\bigr){\rm d}^3{\bf x}.\label{eq11}
\end{eqnarray}
In state $A$ (the actual medium, but with a transparent surface), we choose a unit source for downgoing waves $ q_A^+({\bf x},\omega)=\delta({\bf x}-{\bf x}_S)$
(and $q_A^-({\bf x},\omega)=0$), with ${\bf x}_S$ just above $\partial\mathbb{D}_S$, hence, outside $\mathbb{D}_f$.
Hence, in the fluid we have $p_A^{\pm}({\bf x},{\bf x}_S,\omega)= G_A^{\pm,+}({\bf x},{\bf x}_S,\omega)$, where the second superscript $+$ denotes that the source at ${\bf x}_S$ radiates downward. 
For ${\bf x}$ at  $\partial\mathbb{D}_S$ we have $\partial_3 p_A^+({\bf x},{\bf x}_S,\omega)=\partial_3G_A^{+,+}({\bf x},{\bf x}_S,\omega)=\frac{1}{2}i\omega\rho\delta({\bf x}_{\rm H}-{\bf x}_{{\rm H},S})$ 
(where ${\bf x}_{{\rm H},S}$ denotes the horizontal coordinates of ${\bf x}_S$) and $\partial_3 p_A^-({\bf x},{\bf x}_S,\omega)=\partial_3G_A^{-,+}({\bf x},{\bf x}_S,\omega)=
-\frac{1}{2}i\omega\rho R^\cup({\bf x}_S,{\bf x},\omega)$ (where $R^\cup({\bf x}_S,{\bf x},\omega)$ is the reflection response of the fluid-solid configuration below $\partial\mathbb{D}_S$).
In state $B$ (the actual medium with the free surface), we choose a source for upgoing waves $q_B^-({\bf x},\omega)=\delta({\bf x}-{\bf x}_A)s(\omega)$ (and $q_B^+({\bf x},\omega)=0$),
with ${\bf x}_A$ just above the ocean bottom, hence, inside $\mathbb{D}_f$.
Hence, in the fluid we have $ p_B^{\pm,-}({\bf x},{\bf x}_A,\omega)= G_B^{\pm,-}({\bf x},{\bf x}_A,\omega)s(\omega)$, 
where the second superscript $-$ denotes that the source at ${\bf x}_A$ radiates upward.
Substituting these choices into equation \ref{eq11}, using the pressure-normalized source-receiver reciprocity 
relations $G_B^{+,-}({\bf x},{\bf x}_A,\omega)= G_B^{+,-}({\bf x}_A,{\bf x},\omega)$ and $G_B^{-,-}({\bf x},{\bf x}_A,\omega)= G_B^{+,+}({\bf x}_A,{\bf x},\omega)$ 
 \cite{Wapenaar2020AMP}, yields
\begin{eqnarray}
&& \hspace{-1cm} p_B^{+,+}({\bf x}_A,{\bf x}_S,\omega)- G_A^{+,+}({\bf x}_A,{\bf x}_S,\omega)s(\omega)=\nonumber\\
&&\hspace{0.5cm} \int_{\partial\mathbb{D}_S} R^\cup({\bf x}_S,{\bf x},\omega) p_B^{+,-}({\bf x}_A,{\bf x},\omega) {\rm d}^2{\bf x},\label{eq12}
\end{eqnarray}
where $p_B^{+,\pm}({\bf x}_A,{\bf x},\omega)=G_B^{+,\pm}({\bf x}_A,{\bf x},\omega)s(\omega)$ for all ${\bf x}$ at $\partial\mathbb{D}_S$.
This representation forms an alternative basis for multidimensional down-down deconvolution.
The main difference with equation \ref{eq9} is that here the integral is taken over the sources at ${\bf x}$ at  $\partial\mathbb{D}_S$ instead of over the receivers at $\partial\mathbb{D}_R$,
and that these sources are decomposed into downward and upward radiating components.
Further, note that the downgoing source wavefield $G_A^{+,+}({\bf x}_A,{\bf x}_S,\omega)s(\omega)$ 
does not include the free-surface ghost, unlike the downgoing source wavefield $G_A^+({\bf x}_A,{\bf x}_S,\omega)s(\omega)$ in equation \ref{eq9}.
Apart from notational differences, the representation of equation \ref{eq12} corresponds to those derived by \citet{Boiero2023EAGE} and \citet{Wang2024GEO}.

\begin {figure}[t]
\centerline{\hspace{6.5cm}\epsfysize =8 cm \epsfbox {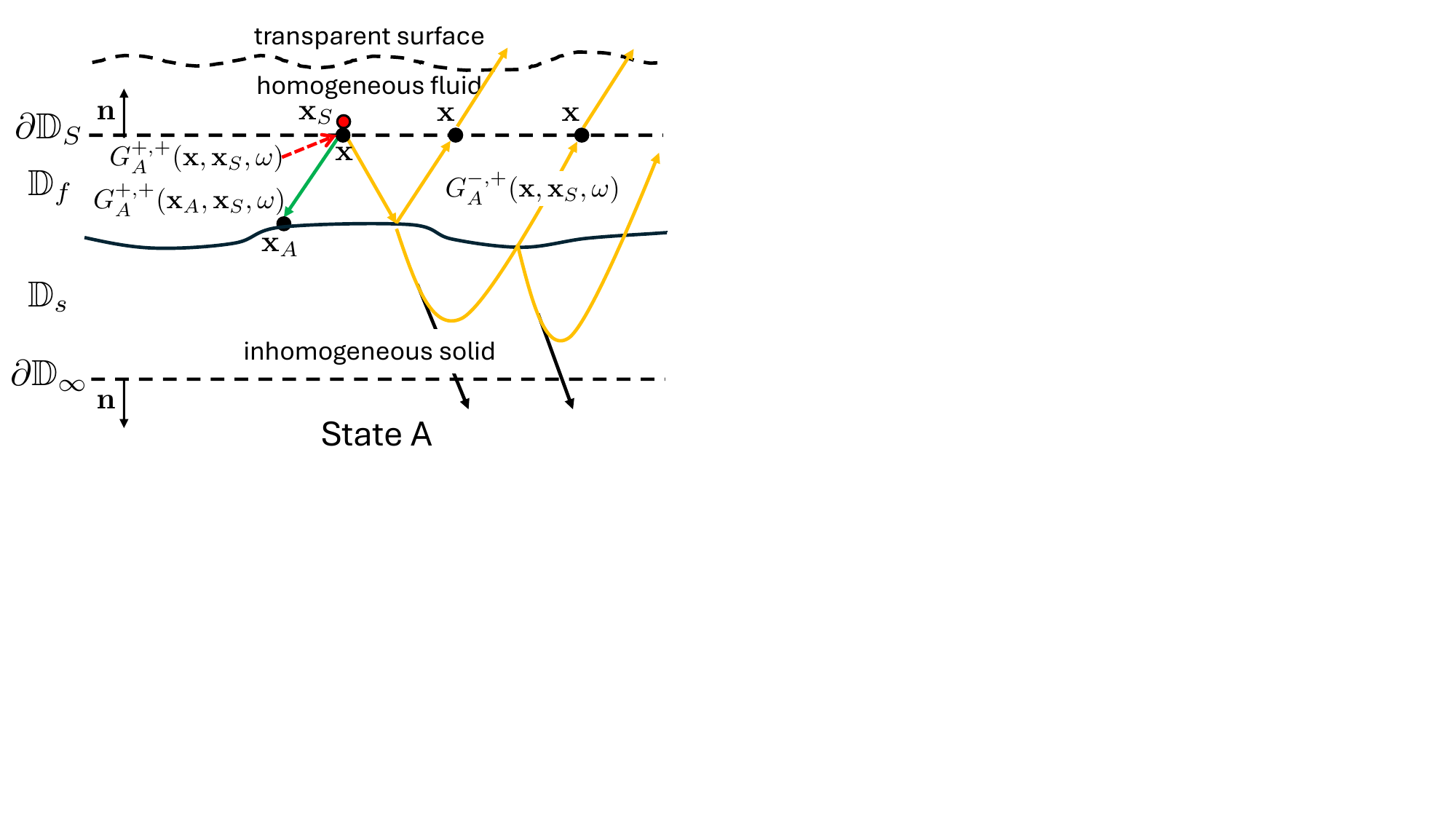}}
\vspace {-2.8cm}
\centerline{\hspace{6.5cm}\epsfysize =8 cm \epsfbox {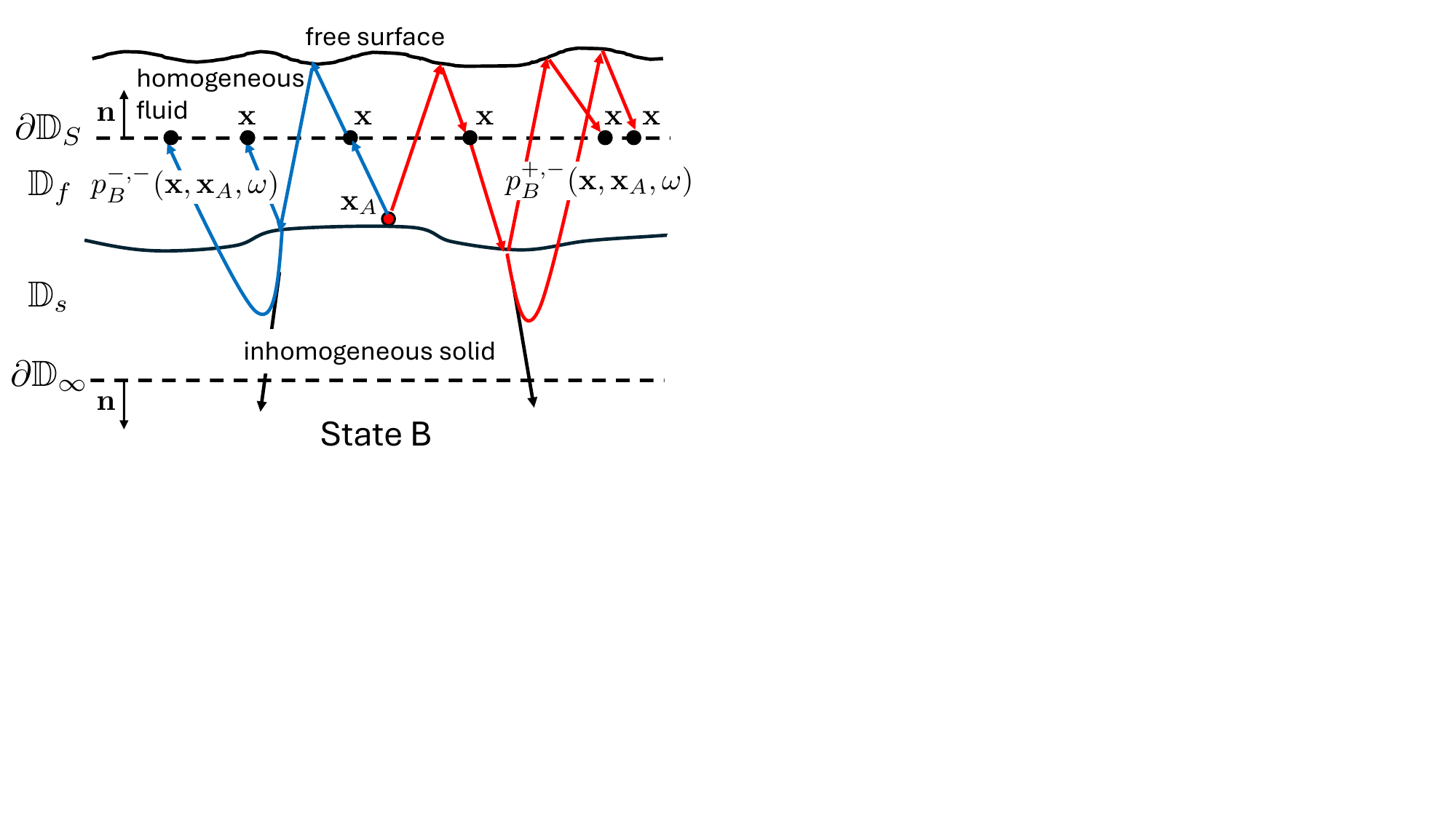}}
\vspace {-3.2cm}
\caption {\small Green's functions, used for the derivation of the source-side multidimensional  down-down deconvolution
representation (equation \ref{eq12}), in an arbitrarily inhomogeneous medium. In state $A$ (the actual medium without the free surface), the Green's function 
$G_A^{+,+}({\bf x}_A,{\bf x}_S,\omega)$ is the downgoing source wavefield (without the free-surface ghost) at ${\bf x}_A$ just above the ocean bottom,
and $G_A^{-,+}({\bf x},{\bf x}_S,\omega)$, with ${\bf x}$ at $\partial\mathbb{D}_S$, represents the reflection response of the fluid-solid configuration, without free-surface multiples.
In state $B$ (the actual medium with the free surface),  $p_B^{\pm,-}({\bf x},{\bf x}_A,\omega)$ represents the downgoing ($+$) and upgoing ($-$) wavefield
at ${\bf x}$ at $\partial\mathbb{D}_S$, in response to an upward radiating source at ${\bf x}_A$.
After applying reciprocity, $p_B^{+,\mp}({\bf x}_A,{\bf x},\omega)$
represents the downgoing wavefield at ${\bf x}_A$ just above the ocean bottom, in response to upward ($-$) and downward ($+$) radiating sources at ${\bf x}$ at $\partial\mathbb{D}_S$.
}\label {Fig3}
\end {figure}

\section{Discussion}

We discuss here the pros ($+$) and cons ($-$) of applying the representations derived in this paper to ocean-bottom data typically acquired for seismic exploration projects. 

Equation \ref{eq8}, i.e. the  representation underlying the traditional
multidimensional up-down  deconvolution method, provides the perfect setup to obtain the receiver-to-receiver response $R^\cup({\bf x}',{\bf x},\omega)$ in a fully data-driven manner.
Given that the integral to be inverted is along the receiver array:\\
$+$\,\, 
it does not require any knowledge of the source locations and wavelets; 
as such it can theoretically handle any complex source signature (including multiple sources firing together - also known as blended data acquisition);\\
$+$\,\, 
 it can handle  complexity in the ``overburden'' 
(e.g. a rough sea surface), making it ideal for 4D compliant processing;\\
$-$\,\, it requires a densely sampled receiver array.

The new representation underlying the receiver-side down-down deconvolution scheme (equation \ref{eq9}), being also formulated as a spatial integral over the receiver array, 
shares most of the pros and cons of equation \ref{eq8}; however, it has some additional pros and cons:\\
$+$\,\, it utilizes only  the downgoing part of the wavefield that better samples the shallow subsurface;\\
$-$\,\, it relies on accurate modeling of the downgoing source wavefield $G_A^+({\bf x}_A,{\bf x}_S,\omega)s(\omega)$ (including the free-surface ghost) 
and of the reflection response $R^\cap({\bf x}_A,{\bf x}',\omega)$ of the water layer from below, which in turn requires knowledge of the acoustic properties of the 
water layer and accurate measurements of the source and receiver positions; in addition, rough sea conditions can further affect the accurate modeling
 of $G_A^+({\bf x}_A,{\bf x}_S,\omega)s(\omega)$ and $R^\cap({\bf x}_A,{\bf x}',\omega)$, which needs to be repeated when the conditions change during a survey.

On the other hand, 
the representation underlying the source-side down-down deconvolution scheme (equation \ref{eq12}) involves an integration over the source array. It has the following pros and cons:\\
$+$\,\,  for economic reasons, the source array is more densely sampled than the receiver array; hence, discretization of 
the source integral in equation \ref{eq12} is more accurate than that of the receiver integrals in equations  \ref{eq8} and  \ref{eq9};\\
$+$\,\, it  accounts for an arbitrarily shaped ocean bottom;\\
$-$\,\, since the integration is carried out over sources, it cannot handle blended data, all source signatures need to be the same, and the sea surface should not change during a survey;\\
$-$\,\, similar to equation \ref{eq9}, since it relies on a model-based downgoing source wavefield $G_A^{+,+}({\bf x}_A,{\bf x}_S,\omega)s(\omega)$, its inversion
is not fully data-driven and therefore less 4D friendly than equation \ref{eq8};\\
$-$\,\, it requires that the downgoing wavefield  $p_B^+({\bf x}_A,{\bf x}_S,\omega)$ is decomposed into a wavefield 
 $p_B^{+,+}({\bf x}_A,{\bf x}_S,\omega)$ due to a source radiating downward and a wavefield $p_B^{+,-}({\bf x}_A,{\bf x}_S,\omega)$  due to a source radiating upward;
in other words, the seismic processing step often referred to as ``source deghosting'' in the seismic exploration community is required;\\
$-$\,\, in the presence of sparsely sampled receiver arrays (or nodes), it represents an underdetermined system of linear equations whose solution requires regularization 
\cite{Boiero2021GEO, Ravasi2025EAGE}.

Finally, it is worth noting that when receiver-side equations \ref{eq8} and \ref{eq9} are approximately solved in a 1-D fashion, this is conventionally accomplished in the receiver gather domain; 
however, our derivation shows that both schemes require dense receiver arrays when treated in a multidimensional fashion. 
As such, whilst the argument of using equation \ref{eq9} over equation \ref{eq8} to leverage the superior illumination of the downgoing wavefield over the upgoing wavefield 
in the presence of sparse receiver geometries is meaningful in 1-D, it falls short in a multidimensional setting due to the need for a dense receiver array in the spatial integral.

\section{Conclusion}

We have derived different representation theorems as bases for suppressing free-surface effects and redatuming  seabed seismic data 
by deconvolving different components of the recorded wavefield. In theory (i.e., under ideal acquisition conditions), the traditional receiver-side up-down 
multidimensional deconvolution scheme is the most appealing in that it is fully data-driven. 
The recently proposed source-side down-down deconvolution scheme represents an important alternative in the presence of sparse receiver geometries 
(when a pre-processing step of receiver interpolation is not an option),  although it requires more careful pre-processing and is no longer fully data-driven. 
On the other hand, our derivation of the representation for multidimensional  receiver-side down-down deconvolution  shows that, 
because of its reliance on a spatial integral over the receiver array, receiver-side down-down deconvolution shares the same challenges as the traditional up-down 
deconvolution method when applied to data acquired with sparse receiver geometries for seismic exploration projects.

\appendix

\section{Appendix A: Reciprocity theorem for a fluid-solid configuration}

Consider the fluid-solid configuration of Figure A--1. It consists of a domain $\mathbb{D}_f$, containing an inhomogeneous fluid, 
and a domain $\mathbb{D}_s$, containing an inhomogeneous solid. These domains are coupled at a fluid-solid interface. 
We define $\partial\mathbb{D}_f$ and $\partial\mathbb{D}_s$ (with outward pointing normal vector ${\bf n}=(n_1,n_2,n_3)$)
as the boundaries of $\mathbb{D}_f$ and $\mathbb{D}_s$, respectively, both excluding the interface.
Hence, the total domain $\mathbb{D}_f\cup\mathbb{D}_s$ is enclosed by  $\partial\mathbb{D}_f\cup\partial\mathbb{D}_s$.
Acoustic and elastodynamic reciprocity theorems hold for domains $\mathbb{D}_f$ and $\mathbb{D}_s$, respectively. Summing the left- and right-hand sides of these theorems,
taking into account that the boundary integrals along the interface cancel each other,  \citet{Hoop90JASA} and \citet{Pandey2025GJI} obtain the following reciprocity theorem for the fluid-solid configuration
\begin{eqnarray}\label{eqA1}
&&\hspace{-.4cm}\int_{\partial\mathbb{D}_f}\bigl( p_A v_{k,B} - v_{k,A} p_B\bigr)n_{k}{\rm d}^2{\bf x}\nonumber\\
&&\hspace{-.4cm}+\int_{\partial\mathbb{D}_s}\bigl( -\tau_{kl,A} v_{k,B} + v_{k,A} \tau_{kl,B}\bigr)n_{l}{\rm d}^2{\bf x}\nonumber\\
&&\hspace{-.4cm}=\int_{\mathbb{D}_f}\bigl(p_A q_B- q_A p_B -v_{k,A}f_{k,B} + f_{k,A}v_{k,B}\bigr){\rm d}^3{\bf x}\nonumber\\
&&\hspace{-.4cm}+\int_{\mathbb{D}_s}\bigl(-\tau_{kl,A} h_{kl,B} +h_{kl,A} \tau_{kl,B}   -v_{k,A}f_{k,B} + f_{k,A}v_{k,B} \bigr){\rm d}^3{\bf x},\nonumber\\
&&\hspace{6.8cm}(\rm A-1)\nonumber
\end{eqnarray}
with subscripts $A$ and $B$ referring again to two different states, 
$p$, $v_k$ and $q$ defined as below equation \ref{eq4}, $\tau_{kl}$ being the stress tensor, $f_k$ the external force density and $h_{kl}$ the deformation-rate density tensor.

\begin {figure}[t]
\centerline{\hspace{7.5cm}\epsfysize =7.5 cm \epsfbox {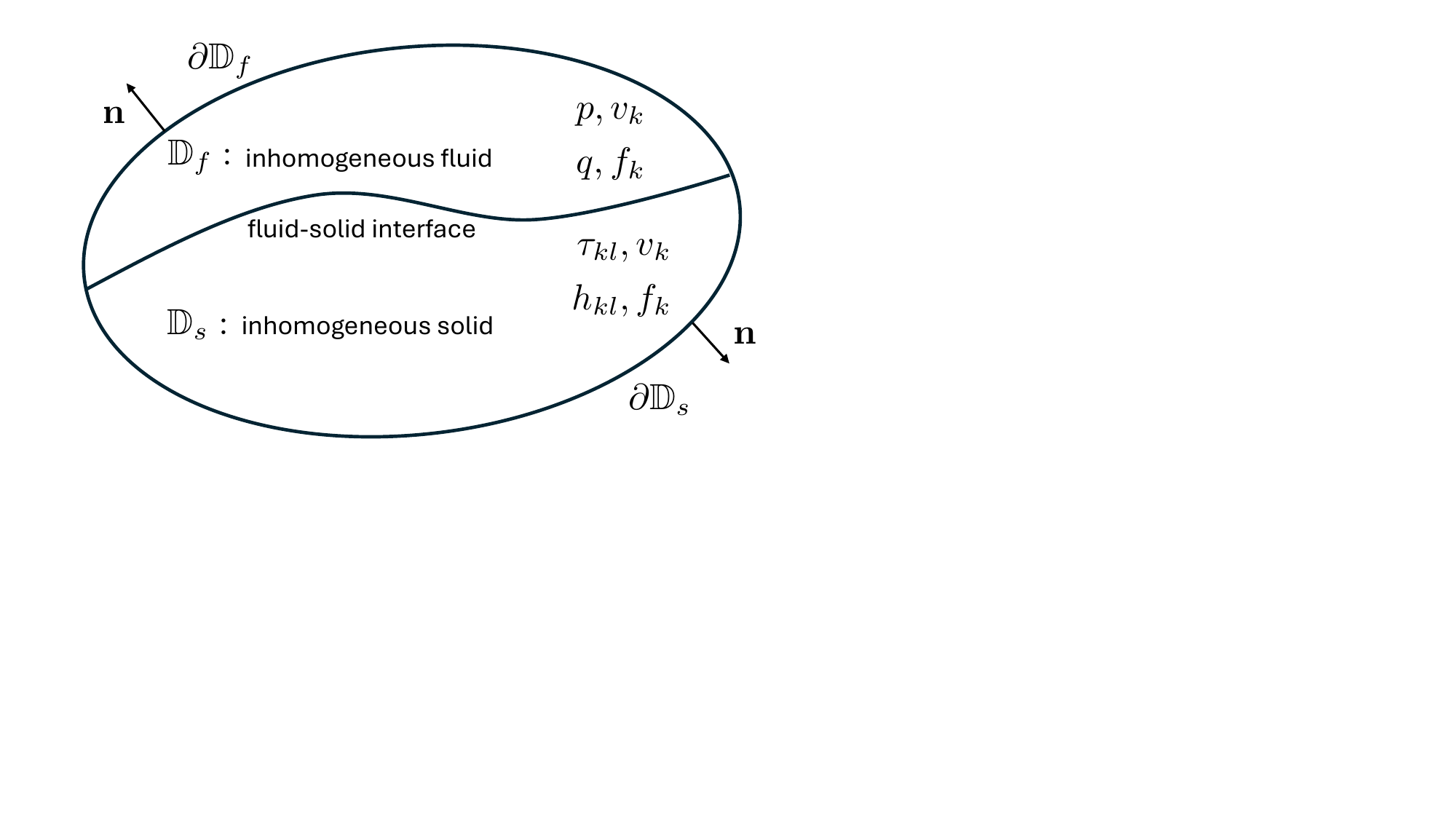}}
\vspace {-3.cm}FIG. A--1 Fluid-solid configuration.
\end {figure}


}

\end{spacing}

\end{document}